\documentclass{svproc}
\usepackage{url}

\begin{document}
\mainmatter              % start of a contribution
\title{Disruptive innovations in RoboCup 2D Soccer Simulation League: from Cyberoos'98 to Gliders2016}
\titlerunning{From Cyberoos'98 to Gliders2016}  % abbreviated title (for running head)
\author{Mikhail Prokopenko$^1$ \and Peter Wang$^{2}$}
\authorrunning{Mikhail Prokopenko and Peter Wang} % abbreviated author list (for running head)
\tocauthor{Mikhail Prokopenko, Peter Wang}
\institute{Complex Systems Research Group, Faculty of Engineering and IT\\ 
			The University of Sydney, NSW 2006, Australia\\
\email{mikhail.prokopenko@sydney.edu.au}\\
\and
Data Mining, CSIRO Data61\\ 
PO Box 76, Epping, NSW 1710, Australia}

\maketitle              % typeset the title of the contribution

\begin{abstract}
We review  disruptive innovations introduced in the RoboCup 2D Soccer Simulation League over the twenty years since its inception, and trace  the progress of our champion team (Gliders). We conjecture that the League has been developing as an ecosystem shaped by diverse approaches taken by participating teams, increasing in its overall complexity. A common feature is that different champion teams succeeded in finding a way to decompose the enormous search-space of possible single- and multi-agent behaviours, by automating the exploration of the problem space with various techniques which accelerated the software development efforts.  These methods included interactive debugging, machine learning, automated planning, and opponent modelling.  The winning approach developed by Gliders is centred on human-based evolutionary computation which optimised several components such as an action-dependent evaluation function, dynamic tactics with Voronoi diagrams, information dynamics, and bio-inspired collective behaviour. 
 
\end{abstract}
\section{Introduction}
\begin{quote}
{\em \small \hspace*{25mm} Agent Smith:   ``You can't win, it's pointless to keep fighting!\\ \hspace*{45mm} Why, Mr. Anderson? Why do you persist?''\\
\hspace*{25mm}  Neo:  ``Because I choose to.''} \vspace{-2mm} \begin{flushright}{\small The Matrix Revolutions.} \end{flushright}
\end{quote}

The first official RoboCup was held in 1997, proposing a new benchmark for Artificial Intelligence (AI) and robotics. Incidentally, another classical AI challenge was successfully met in May 1997 when IBM Deep Blue defeated the human world champion in chess.  By design, RoboCup and chess  differ in a few key elements: environment (static vs dynamic), state change (turn-taking vs real-time), information accessibility (complete vs incomplete), sensor readings (symbolic vs non-symbolic), and control (central vs distributed) \cite{Asada99}.  These differences are emphasised in the RoboCup 2D Soccer Simulation League \cite{Kitano97}, which quickly gained prominence, becoming one of the largest RoboCup leagues. 

In this league, two teams of 12 fully autonomous software programs (called ``agents'') play soccer in a two-dimensional virtual soccer stadium (11 player agents and 1 coach agent in each team), with no remote control. Each player agent receives relative and noisy input from its virtual sensors (visual, acoustic and physical) and may perform some basic actions in order to influence its environment, e.g., running, turning and kicking the ball. The coach agent  receives perfect input but can communicate with the player agents only infrequently and through a fairly limited channel.  The ability to simulate soccer matches without physical robots abstracts away low-level issues such as image processing and motor breakages, allowing teams to focus on the development of complex team behaviours and strategies for a larger number of autonomous agents \cite{rcsymp2014,RAM15}. 

A simulated game lasts just over 10 minutes on average, and is played over a small network of computer workstations which execute the code in parallel.
Each simulation step takes merely a tenth of a second, during which the entire sensory-motor cycle takes place within an agent: starting with receiving new sensory inputs from the simulator, proceeding to updating the internal memory, to evaluating possible choices, to sending the chosen action back to the simulator. The main challenge for each agent is to derive the best possible action to execute at any specific time, while facing unexpected  actions of the opposing agents.

Over 20 years, the RoboCup community has developed the open-source 2D simulator and visualisation software which currently, with various packaged utilities and basic agent libraries, contains nearly a million lines of code. During this period, the League and the participating teams have  undergone several  transitions each of which eventually expanded the level of agents' intelligence and their behavioral complexity.  In this paper we attempt to trace not only the ten-year long progress of our own team from its first implementation (Cyberoos; participated between 1998 and 2003) to the RoboCup-2016 champion team (Gliders; competed first in 2012), but also put this trace in the context of the twenty-year long evolution of the Sim2D League itself. 

The conjecture we put forward is that the League has been developing as an ecosystem with an increasing complexity shaped by different approaches taken by participating teams. Furthermore, this evolving ecosystem has experienced a series of salient transitions leading to  emergence of qualitatively new properties in the intelligence exhibited by the agents. By a transition we do not mean a mere extension of some simulated capabilities, such as the introduction of goalkeepers, heterogeneous player types, or a coach language.  Instead, we associate a transition with a specific methodological advance which played the role of a \emph{disruptive innovation}, with wide-spread consequences affecting the entire ``ecosystem'', for example, a release of standard libraries, and so on. 
We use the term ``disruptive innovation'' in a broad sense to indicate an innovation that creates a new ecosystem (by analogy with a new market or value network), eventually disrupting an existing system, displacing established structures and relationships.

\section{A simulated world}

The foundation supporting the evolution of the League is undoubtedly the construction of the soccer server itself, providing a centralised world model with several key features, enhanced over the following years:
\begin{itemize}
\item distributed  client/server  system  running  on  a  network and producing fragmented,  localised  and  imprecise  (noisy  and latent)  information  about  the  environment  (virtual soccer field) \cite{Noda2003,haker2001method};
\item concurrent  communication  with  a  number of  autonomous agents \cite{Stone99};
\item heterogeneous  sensory  data  (visual,  auditory,  kinetic) without a global  vision, and limited  range  of basic commands/effectors  (turn,  kick,  dash, $\ldots$)  \cite{ATAL2000};
\item asynchronous  perception-action  activity and limited  window of opportunity  to  perform  an  action \cite{Butler2001};
\item autonomous decision-making under  constraints  enforced  by  teamwork (collaboration)  and opponent  (competition) \cite{IAAI2000};
\item conflicts  between reactivity  and deliberation \cite{Reis2001}.
\end{itemize}
The only restriction that was imposed from the outset is  that participants should ``never use central control mechanisms to control a team of agents'' \cite{NodaSMAK98}.

A crucial feature making this simulated world an evolving ``ecosystem'' is the availability of binaries (and sometimes the source code) of participating teams, contained within an online team repository. The repository is updated after each annual RoboCup competition, allowing the participants to improve their teams with respect to the top teams of the previous championships. These improvements diversify the teams' functionality and explore the immense search-space of possible behaviours in the quest for optimal solutions. This process results in a co-evolution of the teams, raising the overall competition level.

\vspace{-2mm}
\section{Partial automation of development efforts}

``AT Humboldt'' from Humboldt University, Germany became the first champion of the League at RoboCup-1997 (Nagoya, Japan). The team used a combination of reactive and planning systems, successfully deploying its agents within the simulated world.  

The following couple of years passed under the domination of ``CMUnited'' team from Carnegie Mellon University (USA) which took the championship in 1998 (Paris, France) and 1999 (Stockholm, Sweden). One of the key reasons for this success was the development of several  tools partially automating the overall effort, such as an offline agent training module, and layered disclosure: a technique for disclosing to a human designer  the specific detailed reasons for an agent actions (in run-time or retroactively).  Layered disclosure made  it  possible  to inspect  the  details  of  an  individual  player's decision-making process at any point \cite{LNAI99-simulator}, becoming in our view the first disruptive innovation in the League.   Together with the offline agent training module, it clearly exemplified the power of automation in accelerating  the development effort --- precisely because it enabled the design effort to reach into a larger part of the search-space by encoding more diverse behaviours.

It is important to point out that there were other novelties introduced by CMUnited-98 and CMUnited-99, such as ``single-channel, low-bandwidth communication'',  ``predictive, locally optimal skills (PLOS)'', ``strategic positioning using attraction and repulsion (SPAR)'', etc. \cite{LNAI99-simulator}, but we believe that it is the partial automation of the software development that became the \emph{disruptive} innovation. It has led to a wide-spread adoption of several debugging, visualising, log-playing, log-analysing, and machine learning tools.

\section{Configurational space}

A number of new teams in 2000 utilised the code base of the 1999 champions, CMUnited-99: it provided code for interaction with the soccer server, skills, strategies, and debugging tools in a variety of programming languages  \cite{StoneABFKLSTW00}.
The champion of RoboCup-2000 held in Melbourne, Australia, ``FC Portugal'' from University of Aveiro and University of Porto, extended this code base with a systematic approach to describing team strategy, the concepts of tactics, formations and player types, as well as the situation based strategic positioning, the dynamic positioning and role exchange mechanisms \cite{ReisL00,Reis2001}.

The generic innovation underlying these mechanisms comprised the ability to configure diverse single- and multi-agent behaviours. The range of these behaviours span from active (ball possession) to strategic (ball recovery), from formations to tactics, and from individual skills to team strategies.  Such diversity resulted in a considerable configurational flexibility displayed by the winning team, significantly increasing the software development productivity, and more importantly, expanding the extent of the available behavioural search-space.  

Not surprisingly, the expansion brought about by the larger configurational capacity was further exploited by the introduction of a standard coach language \cite{P-008-M65} enabling high-level coaching with explicit definition of formations, situations, player types and time periods, and resulting in a high-level coordination of team behaviour.  In other words, a disruptive innovation again was delivered by a method which allowed to access deeper regions of the available search-space.

Team ``TsinghuAeolus'' from Tsinghua University, China, which won the next two championships (RoboCup-2001 in Seattle, USA, and RoboCup-2002 in Fukuoka), focussed specifically on increasing the agents' adaptability via a novel online advice-taking mechanism \cite{Jinyi2004}. The configurational space was extended by a task-decomposition mechanism that assigned different parts of the task to different agents. 

A major boost to the League was provided by the partial release of the source code of the next champion, team ``UvA Trilearn'' from University of Amsterdam, The Netherlands, which won RoboCup-2003 in Padua, Italy \cite{Kok03robocup}.  This release resulted in a standardisation of many low-level behaviours and world model, effectively ``locking in'' the configurational space attained by that time, and motivating several teams to switch their code base to UvA Trilearn base.

\section{Cyberoos: 1998 -- 2003}	

At this stage we take a brief look at our first team, Cyberoos, which participated in RoboCup competitions between 1998 and 2003. The Cyberoos'98 team took
$3^{rd}$ place in the 1998 Pacific Rim RoboCup competition \cite{prokopenko1998designing}, while Cyberoos'2000 were $4^{th}$ in the Open European RoboCup-2000 \cite{Butler2001}. Despite these regional successes, the team's best result at the world stage was a shared $9^{th}$ place which Cyberoos repeatedly took at the RoboCup competitions in 2000, 2001, 2002 and 2003, never reaching the quarter-finals \cite{Prokopenko2001,ProkopenkoWH01,Prokopenko02,Prokopenko03}. In hindsight, the main reason for this lack of progress was an oversight of the main tendency driving the innovations in the League: the exploration of the search-space due to the automation of the development efforts and the standardisation of the configurational space. 

Instead, the approach taken by Cyberoos focussed on self-organisation of emergent behaviour within a purely reactive agent architecture \cite{ProkopenkoWH01}.  Only during the later years the Cyberoos architecture diversified, and included semi-automated methods that quantified the team performance in generic information-theoretic terms  \cite{Prokopenko02,Prokopenko03}. This approach focussed on measuring the behavioural and belief dynamics in multi-agent systems, offering a possibility to evolve the team behaviour,  optimised under a universal objective function, within the framework of information-driven self-organisation \cite{nehaniv2005evolutionary,prokopenko06-alife,sab06}. However, this framework has started to take a functional shape only a few years later, after the time when the Cyberoos team effort stopped in 2003. 

\vspace{-3mm}
\section{Search-space decomposition}
\vspace{-1mm}

The next decade of RoboCup championships witnessed an intense competition between three teams: ``Brainstormers'' from University of Osnabr{\"u}ck, Germany,  ``WrightEagle'' from University of Science and Technology of China, and ``HELIOS'' from Fukuoka University and Osaka Prefecture University, Japan.   Brainstormers became champions three times: in 2005 (Osaka, Japan), 2007 (Atlanta, USA), and 2008 (Suzhou, China);  WrightEagle came first an incredible six times: in 2006 (Bremen, Germany), 2009 (Graz, Austria), 2011 (Istanbul, Turkey), 2013 (Eindhoven, The Netherlands), 2014 (Joao Pessoa, Brazil) and 2015 (Hefei, China); and HELIOS succeeded twice: in 2010 (Singapore) and 2012 (Mexico City, Mexico). 

\vspace{-2mm}
\subsection{Machine learning}

Brainstormers' effort focussed on reinforcement learning methods aiming at a universal machine learning system, where the agents learn to generate the appropriate behaviors to satisfy the most general objective of ``winning the match''. Unfortunately, as has been acknowledged \cite{gabel2008brainstormers}, ``even from very optimistic complexity estimations it becomes obvious, that in the soccer simulation domain, both conventional solution methods and also advanced today's reinforcement learning techniques come to their limit --– there are more than $(108 \times 50)^{23}$ different states and more than $(1000)^{300}$ different policies per agent per half time''.  

The high dimensionality of the search space motivated Brainstormers to use a multilayer perceptron neural network \cite{gabel2008brainstormers}: a feedforward artificial neural network which utilises a supervised learning technique called backpropagation for training the network. Rather than developing a universal learning system, Brainstormers succeeded in decomposing the problem into a number of individual behaviours (e.g., NeuroKick, NeuroIntercept, NeuroHassle) and tactics (e.g., NeuroAttack2vs2, NeuroAttack3vs4, NeuroAttack7vs8), learned with supervised learning techniques.  

Recently, there has been some renewed interest in backpropagation networks due to the successes of deep learning.  In our view, the potential of reinforcement learning methods in RoboCup has not yet been fully realised, and deep learning may yet to become a disruptive innovation for the Simulation league.

\subsection{Automated planning}

WrightEagle team addressed the challenges of (i) high dimensionality of the search space and (ii) the limited computation time available in each decision cycle, by using Markov Decision Processes (MDPs). The developed framework decomposes a given MDP into a set of sub-MDPs arranged over a hierarchical structure, and includes heuristics approximating online planning techniques \cite{zhang2013decision}. WrightEagle approach abandoned ``the pursuit of absolute accuracy'' and divided the continuous soccer field into the discrete space, further subdividing it into the players' control areas according to geometric reachability.  The resultant structure enables automated planning, accelerating the search process and extending the search depth \cite{zhang2013decision}.

\vspace{-2mm}
\subsection{Opponent modelling}

``HELIOS'' team \cite{Akiyama2008,Akiyama2010} followed a similar path, targeting a decomposition of the problem space in developing an unsupervised learning method based on Constrained Delaunay Triangulation (CDT) \cite{chew1989constrained}. A Delaunay triangulation for a set $P$ of points in a plane is a triangulation $\cal{D}(P)$ such that no point in $P$ is inside the circumcircle of any triangle in $\cal{D}(P)$ (in CDT the circumcircle of some triangles  contains other triangles' vertices).  The method divides the soccer field into a set of triangles, which provide an input plane region for Neural Gas (NG) and Growing Neural Gas (GNG) methods. Specifically, the set $P_b$ of $N$ points represents specifically chosen positions of the ball on the field, while sets $P_i$ describe the sets of coordinates of each player $1 \le i \le 11$, so that there is a bijective correspondence between $P_b$ and each of $P_i$. Moreover, when the ball takes any position within a triangle of ${\cal{D}}(P_b)$, each player's position is computed in a congruent way within ${\cal{D}}(P_i)$.   During offline experiments or even during a game, the behaviour of the opponent, for example,  the players' motion, directions of the passes, and the overall team formations, can be mapped, analysed and categorised \cite{Akiyama2008,Akiyama2010}.

It is evident that the main reason behind the recurrent successes of all three champion approaches is a dynamic decomposition of the problem space  and its subsequent efficient exploration. This innovation goes beyond a simple standardisation of  low-level behaviours within a rich but static configurational space, by employing automated learning and planning methods in a dynamic search.

\section{Standardisation of ``hardware''}

An influential disruptive innovation arrived in 2010, when HELIOS team released a major update of their well-developed code base \cite{agent2d}:
\begin{itemize}
\item \emph{librcsc-4.0.0}: a base library for the RoboCup Soccer Simulator (RCSS);
\item \emph{agent2d-3.0.0}: a base source code for a team;
\item \emph{soccerwindow2-5.0.0}: a viewer and a visual debugger program for RCSS;
\item \emph{fedit2-2.0.0}: a team formation editor for \emph{agent2d}.
\end{itemize}
This resulted in nearly 80\% of the League's teams switching their code  base to agent2d over the next few years.  One may think of this phenomenon as a standardisation of the simulated hardware, freeing the effort to improving the higher-level tactical behaviours.

\section{Gliders (2012 -- 2016): Fusing human innovation and artificial evolution}

We turn our attention to our champion team which won  RoboCup-2016 (Leipzig, Germany):  Gliders \cite{POWH12,gliders2013tdp,gliders2014tdp,gliders2015tdp,gliders2016tdp}.  Gliders2012 and Gliders2013 reached the semi-finals of RoboCup in 2012 and 2013;  Gliders2014 were runner-ups in 2014; Gliders2015 finished third in RoboCup-2015, and Gliders2016 (a joint effort of the University of Sydney and CSIRO) became world champions in 2016. 

RoboCup-2016 competition  included 18 teams from 9 countries: Australia, Brazil, China, Egypt,
Germany, Iran, Japan, Portugal and Romania. Gliders2016 played 23 games during several rounds, winning 19 times, losing twice and drawing twice, with the total score of 62:13, or 2.70 : 0.57 on average. In the two-game semi-final round, Gliders2016 defeated team CSU\_Yunlu from Central South University (China),  winning both games with the same score 2:1.  The single-game final against team HELIOS2016 (Japan) went into the extra time, and ended with Gliders2016 winning 2:1.  The third place was taken by team Ri-one from Ritsumeikan University (Japan).

The 2016 competition also included an evaluation round, where all 18 participating teams played one game each against the champion of RoboCup-2015, team WrightEagle (China).  Only two teams, the eventual finalists Gliders2016 and HELIOS2016, managed to win against the previous year champion, with Gliders defeating WrightEagle 1:0, and HELIOS producing the top score 2:1. 

The Gliders team code is written in C++ using \emph{agent2d-3.1.1} \cite{agent2d}, and fragments of source code of team MarliK released in 2012 \cite{marlik}. 

In order to optimise the code, the Gliders development effort over the last five years involved \emph{human-based evolutionary computation} (HBEC): a set of evolutionary computation techniques that rely on human innovation \cite{kosorukoff2001human,Cheng2004}. 

In general, evolutionary algorithms search a large space of possible solutions that together form a population. Each solution is a ``genotype'': a complex data structure representing the entire team behaviour encoded through a set of ``design points''. A design point can be as simple as a single parameter  (e.g., risk tolerance  in making a pass), or as complicated as a multi-agent tactical behaviour (e.g., a conditional statement describing the situation when a defender moves forward to produce an offside trap). 

Some design points are easy to vary. For instance, a formation defined via Delaunay Triangulations ${\cal{D}}(P_b)$ and ${\cal{D}}(P_i)$, $1 \le i \le 11$, is an ordered list of coordinates, and varying and recombining such a list can be relatively easily automated.  Other design points have an internal structure and are harder to permute.  For example, a conditional statement describing a tactic has a condition and an action, encoded by numerous parameters such as positional coordinates, state information, and action details. Once such a statement (a design point) is created by human designers, its encoding can be used by evolutionary algorithms. However, the inception of the tactic needs creative innovation in the first place, justifying the hybrid HBEC approach.      

The HBEC solutions representing team behaviours are evaluated with respect to their fitness, implemented as the average team performance, estimated over thousands of games for each generation played against a specific opponent. Some solutions are retained and recombined (i.e, the members of the population live) and some are removed (i.e., die) through selection.  Importantly, the evolutionary process is carried out within different landscapes (one per known opponent), and typically results in different solutions evolved to outperform specific opponents. In order to maintain coherence of the resultant code, each design point is implemented with a logical mask switching the corresponding part of the genotype on and off for specific opponents (determined by their team names). This is loosely analogous to epigenetic programming \cite{Tanev2008}.     

The approach is aimed at constantly improving performance from one artificial ``generation" to another, with team designers innovating and recombining behaviours while the fitness landscape and the mutations are for the most part automated. The performance of Gliders was evaluated on several supercomputer clusters, executing on some days tens of thousands of the experimental runs with different behaviour versions. It would be a fair estimate that the number of such trials is approaching 10 million.   The overall search-space explored by the HBEC includes variations in both Gliders behaviour and opponent modelling. The approach incorporates disruptive innovations of the past years, including the standardisation of simulated ``hardware'' and several effective search-space decompositions.
  
Specific variations included (i) action-dependent evaluation function, (ii) dynamic tactics with Voronoi diagrams, (iii) information dynamics, and (iv) bio-inspired collective behaviour.

The approach introduced in Gliders2012 \cite{POWH12} retained the advantages of a single evaluation metric (implemented in \emph{agent2d}  \cite{agent2d}), but diversified the evaluation by considering multiple points as desirable states.  These desirable states for action-dependent evaluation are computed using Voronoi diagrams which underlie many tactical schemes of Gliders. 

Starting from 2013, Gliders utilised information dynamics \cite{liz10e,Wang12,ay-robot,liz10d,cliff2013towards,liz08b} for tactical analysis and opponent modelling.  This analysis involves computation of information transfer and storage, relating the information transfer  to responsiveness of the players,  and the information storage within a team to the team's rigidity and lack of tactical richness.

The constraints on mobility, identified by the information dynamics, were investigated and partially overcome with bio-inspired  collective behaviour \cite{gliders2015tdp}.  Gliders2015 utilise several elements of swarm behavior, attempting to keep each player's position as close as possible to that suggested by a specific tactical scheme, while incorporating slight variations in order to maximise the chances of receiving the pass and/or shooting at the opponent's goal.   This behaviour increased the degree of coherent mobility: on the one hand, the players are constantly refining their positions in response to opponent players, but on the other hand, the repositioning is not erratic and the players move in coordinated ways.    

These directions were unified within a single development and evaluation framework which allowed to explore the search-space in two ways: translating human expertise into new behaviours and tactics, and exhaustively recombining them with an artificial evolution, leveraging the power of modern supercomputing. This fusion, we believe, produced a disruptive innovation on its own, providing the winning edge for Gliders.

\section{Conclusion}			
\vspace{-1mm}

In this paper we reviewed  disruptive innovations which affected advancement of the RoboCup 2D Soccer Simulation  League over the twenty years since its inception, and placed the progress of our champion team in this context. It is important to realise that the neither of these processes has been linear, and many ideas have been developing  along a spiral-shaped trajectory, resurfacing over the years in a different implementation. For example, the utility of evolutionary computation supported by supercomputing has been suggested as early as 1997, when a simulated team was developed with the agents whose high-level decision making
behaviors had been entirely evolved using genetic programming \cite{Luke98geneticprogramming}. Yet the complexity of the domain proved to be too challenging for this approach to gain a widespread adoption at that time.  

Without an exception, all the winning approaches combined elements of some automation (debugging, machine learning, planning, opponent modelling, evolutionary computation) with human-based innovation in terms of a decomposition of the search-space, providing various configurations, templates and structures. Is there still a way toward a fully automated solution, when the agents learn or evolve to play a competitive game without a detailed guidance from human designers, but rather by trying to satisfy a universal objective  (``win a game'')?

On the one hand, the ability to run a massive number of simulated games on supercomputing clusters producing replicable results will only strengthen in time \cite{RAM15}, and so may lend some hope in meeting this challenge positively. On the other hand, the enormous size and dimensionality of the search-space  would defy any unstructured exploration strategy.  A methodology successfully resolving this dilemma may not only provide an ultimate disruptive innovation  in the League, but also provide a major breakthrough in the general AI research.

\vspace{-3mm}
\section*{Acknowledgments}
\vspace{-2mm}

Several people contributed to Cyberoos and Gliders development over the years. Marc Butler, Thomas Howard and Ryszard Kowalczyk made exceptionally valuable  contributions to Cyberoos' effort during 1998 -- 2002 \cite{prokopenko1998designing,Butler2001,Prokopenko2001,ProkopenkoWH01}. We are grateful to Gliders team members Oliver Obst, particularly for establishing the tournament infrastructure supporting the team's performance evaluation on CSIRO Accelerator Cluster (Bragg), and Victor Jauregui, for several important insights on soccer tactics used in Gliders2016 \cite{gliders2016tdp}.  We thank David Budden for developing a new self-localisation method introduced in Gliders2013 \cite{budden2013particle,gliders2013tdp} as well as contributing to the analysis of competition formats \cite{RAM15}, and Oliver Cliff for developing a new communication scheme adopted by Gliders from 2014 \cite{gliders2014tdp}.  The overall effort has also benefited from the study quantifying tactical interaction networks, carried out in collaboration with Oliver Cliff, Joseph T. Lizier, X. Rosalind Wang and Oliver Obst \cite{cliff2013towards}. We are thankful to Ivan Duong, Edward Moore and Jason Held for their contribution to Gliders2012 \cite{POWH12}.  Gliders team logo was created by Matthew Chadwick.


\bibliographystyle{splncs}

\end{document}